\documentclass[10pt,twocolumn,letterpaper]{article}

\usepackage{iccv}
\usepackage{times}
\usepackage{epsfig}
\usepackage{graphicx}
\usepackage{amsmath}
\usepackage{amssymb}
\usepackage{algorithm}
\usepackage[noend]{algpseudocode}
\usepackage[breaklinks=true,bookmarks=false]{hyperref}

\iccvfinalcopy % *** Uncomment this line for the final submission

\def\iccvPaperID{****} % *** Enter the ICCV Paper ID here

\ificcvfinal\fi

\begin{document}

%%%%%%%%% TITLE
\title{Deep Plug-and-Play Prior for Parallel MRI Reconstruction}

\author{Ali Pour Yazdanpanah\textsuperscript{1,2} \ \ Onur Afacan\textsuperscript{1,2} \ \ Simon K. Warfield\textsuperscript{1,2}\\
\textsuperscript{1}Harvard Medical School\\
\textsuperscript{2}Boston Children's Hospital\\
{\tt\small \{ali.pouryazdanpanahkermani,onur.afacan,simon.warfield\}@childrens.harvard.edu}
}

\maketitle
\ificcvfinal\thispagestyle{empty}\fi

%%%%%%%%% ABSTRACT
\begin{abstract}
Fast data acquisition in Magnetic Resonance Imaging (MRI) is vastly in demand and scan time directly depends on the number of acquired k-space samples. Conventional MRI reconstruction methods for fast MRI acquisition mostly relied on different regularizers which represent analytical models of sparsity. However, recent data-driven methods based on deep learning has resulted in promising improvements in image reconstruction algorithms. In this paper, we propose a deep plug-and-play prior framework for parallel MRI reconstruction problems which utilize a deep neural network (DNN) as an advanced denoiser within an iterative method. This, in turn, enables rapid acquisition of MR images with improved image quality. The proposed method was compared with the reconstructions using the clinical gold standard GRAPPA method. Our results with undersampled data demonstrate that our method can deliver considerably higher quality images at high acceleration factors in comparison to clinical gold standard method for MRI reconstructions. Our proposed reconstruction enables an increase in acceleration factor, and a reduction in acquisition time while maintaining high image quality.
\end{abstract}

%%%%%%%%% BODY TEXT
\section{Introduction}

Parallel imaging (PI) techniques have become popular strategies for reducing scan time, such as by undersampling phase encodes in different MRI sequences, and for mitigating geometric distortion, such as in EPI sequences. PI uses spatially varying coil sensitivity profiles from an array of receiver coils, to reconstruct images despite undersampling of data. Two primary PI reconstruction algorithms are the image domain method of sensitivity encoding (SENSE) \cite{pruessmann1999sense} and the frequency domain method of generalized autocalibrating partially parallel acquisitions (GRAPPA) \cite{griswold2002generalized}. SENSE reconstruction is achieved by solving a linear system of equations that uses the coil sensitivity profiles. Efficient reconstruction methods have been introduced based on the SENSE framework for both Cartesian \cite{pruessmann1999sense}, and non-Cartesian trajectories \cite{pruessmann2001advances}. SENSE-based methods, that explicitly require the coil array sensitivity maps to be known, include parallel MRI with adaptive radius in $k$-space (PARS) \cite{yeh20053parallel}, parallel imaging reconstruction for arbitrary trajectories using $k$-space sparse matrices (kSPA) \cite{liu2007parallel}, and sensitivity profiles from an array of coils for encoding and reconstruction in parallel (SPACE-RIP) \cite{kyriakos2000sensitivity}. The PARS approach performs point-by-point reconstruction by creating and inverting small localized encoding matrices (lies within a small radius in $k$-space) instead of inverting the complete encoding matrix at once.
The kSPA approach performs the reconstruction as a system of sparse linear equations in k-space which is achieved by computing a sparse approximate inverse matrix.
The SPACE-RIP approach positions RF receiver coils around the object of interest which make it possible to image different plane within the volume of interest and also enabling variable density sampling in $k$-space.\\
GRAPPA reconstruction is an autocalibrating method based on local $k$-space kernels, which utilize the learned correlation between multiple channels in local areas and fill missing k-space values by a linear combination of the acquired local data from multiple coils. Partially parallel imaging with localized sensitivities (PILS) \cite{griswold2000partially}, simultaneous acquisition of spatial harmonics (SMASH) \cite{sodickson1997simultaneous}, AUTO-SMASH \cite{jakob1998auto}, and VD-AUTO-SMASH \cite{heidemann2001vd}, are other examples of autocalibrating methods which utilize $k$-space kernels. The PILS approach needs specific coil arrangement in order to reconstruct the full field of view. The receiver coils in PILS approach located linearly in the direction of phase encoding with localized sensitivities over different regions in the full field of view.
The SMASH approach, generates missing data directly by a weighted linear combination of the estimated sensitivity maps. The weights in SMASH are estimated by fitting the maps to spatial harmonic of specific order.
The AUTO-SMASH approach exploits a few numbers of autocalibration signal (ACS) lines to estimate the sensitivities. Variable-density AUTO-SMASH (VD-AUTO-SMASH) uses multiple ACS lines from the $k$-space center to improve the reconstruction of the AUTO-SMASH approach. GRAPPA approach can be considered as a generalized version of the VD-AUTO-SMASH approach. GRAPPA techniques have been extended to non-Cartesian $k$-space trajectories by the methods such as \cite{seiberlich2008reconstruction,heberlein2006auto}. Iterative self-consistent parallel imaging reconstruction (SPIRIT) \cite{lustig2010spirit} method is also based on GRAPPA framework but formulated as an inverse problem which can reconstruct data from arbitrary $k$-space trajectories. The extended version of SPIRIT method was introduced in \cite{uecker2014espirit} (ESPIRIT) which considered to reconstruct missing data by restricting the solution to a subspace and estimate the sensitivity maps from autocalibration lines in $k$-space using eigenvalue decomposition of $k$-space filtered calibrated kernels in image space.

It is also possible to reduce the aliasing artifact through techniques such as 2D CAIPIRINHA \cite{breuer2006controlled}, and Wave-CAIPI \cite{bilgic2015wave}, both of which modify the data acquisition in order to reduce the concentration of aliasing artifacts on certain regions, leading to a reduced geometry factor penalty, and a more robust reconstruction.

A popular approach to reduce or mitigate the presence of reconstruction artifacts has been regularized reconstruction or compressed sensing (CS) method. Methods based on SENSE are particularly reformulated to include regularization \cite{ramani2011parallel}, and rapid and accurate reconstructions are possible. Popular regularization constraints include gradient norm, total variation, and $l_1$-norm of transform domain coefficients to promote edge preservation and sparsity. CS methods seek to exploit intrinsic image properties of sparsity in a transform domain and have allowed for highly accelerated imaging in different settings. These techniques allow for images to be reconstructed with the similar linear algebra equation that express both data similarity term and sparsity penalty term. These have allowed to increase data acquisition speed while generating better reconstructions \cite{ramani2011parallel,lustig2007sparse,liang2009accelerating,bammer2007augmented,guerquin2012realistic,yazdanpanah2017compressed,ayazdanpanah2017,haldar2014low}.
Due to nondifferentiability of some regularizers, proximal methods like alternating direction method of multipliers (ADMM) \cite{boyd2011distributed} has been proposed. 
Proximal methods uses proximal operator to avoid the regularizer differentiation. The plug-and-play prior framework \cite{venkatakrishnan2013plug} presented with an idea to utilize the denoiser without any regularization objective as proximal operator in an iterative method for image recovery. The method has been used in different imaging inverse problem applications \cite{sreehari2016plug,chan2017plug,kamilov2017plug,sun2019online}. In \cite{sreehari2016plug}, authors used the plug-and-play framework for bright field electron tomography. In \cite{chan2017plug}, plug-and-play alternating direction method of multipliers has been used for image restoration applications. In \cite{kamilov2017plug}, the authors developed the fast-iterative shrinkage/thresholding algorithm (FISTA) variant of plug-and-play prior for model-based nonlinear inverse scattering and proved that the framework is applicable beyond linear inverse problems. In \cite{sun2019online}, the authors introduced a scalable version of plug-and-play framework based on iterative shrinkage/thresholding algorithm (ISTA) which utilized a subset of measurement at every iteration in order to parallelize the algorithm. In all the mentioned papers, a fixed denoiser has been used as the proximal operator which its accuracy can't be ideal in different scenarios for different applications. However, in this paper we present a learning-based plug-and-play prior framework for parallel MRI reconstruction which extends the framework to its data-adaptive variant and provides an end-to-end reconstruction scheme. We evaluate the reconstruction performance of our method to clinically-used GRAPPA method. GRAPPA reconstruction method is being considered as the clinical gold standard and the most trusted method by radiologist for MRI reconstruction.
\\
\section{Methods}

The complete MR imaging model given by

\begin{equation}\label{Eq1}
\text d_l(k_m)=\int \text S_l(\rho) \text x(\rho) e^{-i2\pi k_m\rho} d\rho + n_l(k_m).
\end{equation}

where $\text d_l(k_m)$ is the data-samples measurements from $l_{th}$ coil at the $m_{th}$ $k$-space location $k_m$. $n_l(k_m)$ is the noise measured from $l_{th}$ coil at the $m_{th}$ $k$-space location. $\text x(\rho)$ is the samples of unknown MR image to be recovered. $\text S_{l}$ is the sensitivity map of the $l_{th}$ coil. The following MR imaging model is the discretized version of Equation (\ref{Eq1}):

\begin{equation}\label{Eq2}
\text d=\text E \text x + \text n.
\end{equation}

where $\text x$ is the samples of unknown MR image, $\text E =\text {PFS}$ is an encoding matrix, and $\text F$ is a Fourier matrix. $\text P$ is a mask representing k-space undersampling pattern and $\text S=[\text S_{1}...\text S_{L}]$, $\text S_{l}$ is a matrix representing the sensitivity map of the $l_{th}$ coil, $1 \leq l \leq L$, and $L$ is the total number of coils. 

Assuming without loss of generality that the inter-coil noise covariance has been whitened, the imaging model can be solved and reach the optimal maximum likelihood estimate for $\text x$ when $E$ has full column rank. This can be done by solving the following least squares problem 

\begin{equation} \label{Eq3}
\begin{aligned}
\hat{\text x} =\underset{\text x}{ argmin} \ \frac{1}{2}\|\text d-\text E\text x\|_{2}^{2}
\end{aligned}
\end{equation}

which results in

\begin{equation} \label{Eq4}
\begin{aligned}
\hat{\text x} =(\text E^{H}\text E)^{-1}\text E^{H} \text d
\end{aligned}
\end{equation}

In a case of undersampled $k$-space data, Equation (\ref{Eq4}) yields artifacts depending on the sampling pattern. 
If we consider Cartesian-type sampled $k$-space, then we create aliasing artifacts in the coil images. 

Equation (\ref{Eq3}) can be poorly-conditioned in a case of high acceleration factor, and
therefore, a regularization term can be incorporated additionally in the least-square approach. Assuming that the interchannel noise covariance has been whitened, the reconstruction relies on the regularized least-square approach:

\begin{equation} \label{Eq5}
\begin{aligned}
\hat{\text x} =\underset{\text x}{ argmin} \ \frac{1}{2}\|\text d-\text {PFS}\text x\|_{2}^{2}+\beta \text R(\text x)
\end{aligned}
\end{equation}

where $\text R$ is a regularization functional that promotes sparsity in the solution and $\beta > 0$ controls the intensity of the regularization.

Our iterative deep plug-and-play prior framework based on ADMM for solving the Equation (\ref{Eq5}) is provided in Algorithm \ref{alg1}.

\begin{algorithm}
\caption{Deep Plug-and-Play Prior}\label{alg1}
\textbf{Input:} $\text x^{0},\text d,\text S,\text u^{0}=0, \lambda>0$
\begin{algorithmic}[1]
\For {$i=1,2,...,N$}
\State $a^i\gets prox(\text d,\text S, \text x^{i-1}-\text u^{i-1};\lambda)$
\State $\text x^i\gets \text{DNN}(a^i+\text u^{i-1})$
\State $\text u^i\gets u^{i-1} + (a^i-\text x^i)$
\EndFor\label{}
\State \textbf{end for}
\end{algorithmic}
\end{algorithm}

$\text x^{0}=\text E^{H}\text d$ is used as an initialization to the algorithm. For least-square cases, we have

\begin{equation} \label{Eq6}
\begin{aligned}
prox (\text d,\text S, \widetilde{\text x};\lambda) = \underset{\text z}{ argmin} \ \frac{1}{2}\|\text z-\widetilde{\text x}\|_{2}^{2}+ \frac{\lambda}{2}\|\text {PFS}\text z-\text d\|_{2}^{2}
\end{aligned}
\end{equation}

\begin{figure*}
\begin{center}
\centerline{\includegraphics[width=\linewidth,scale=0.6]{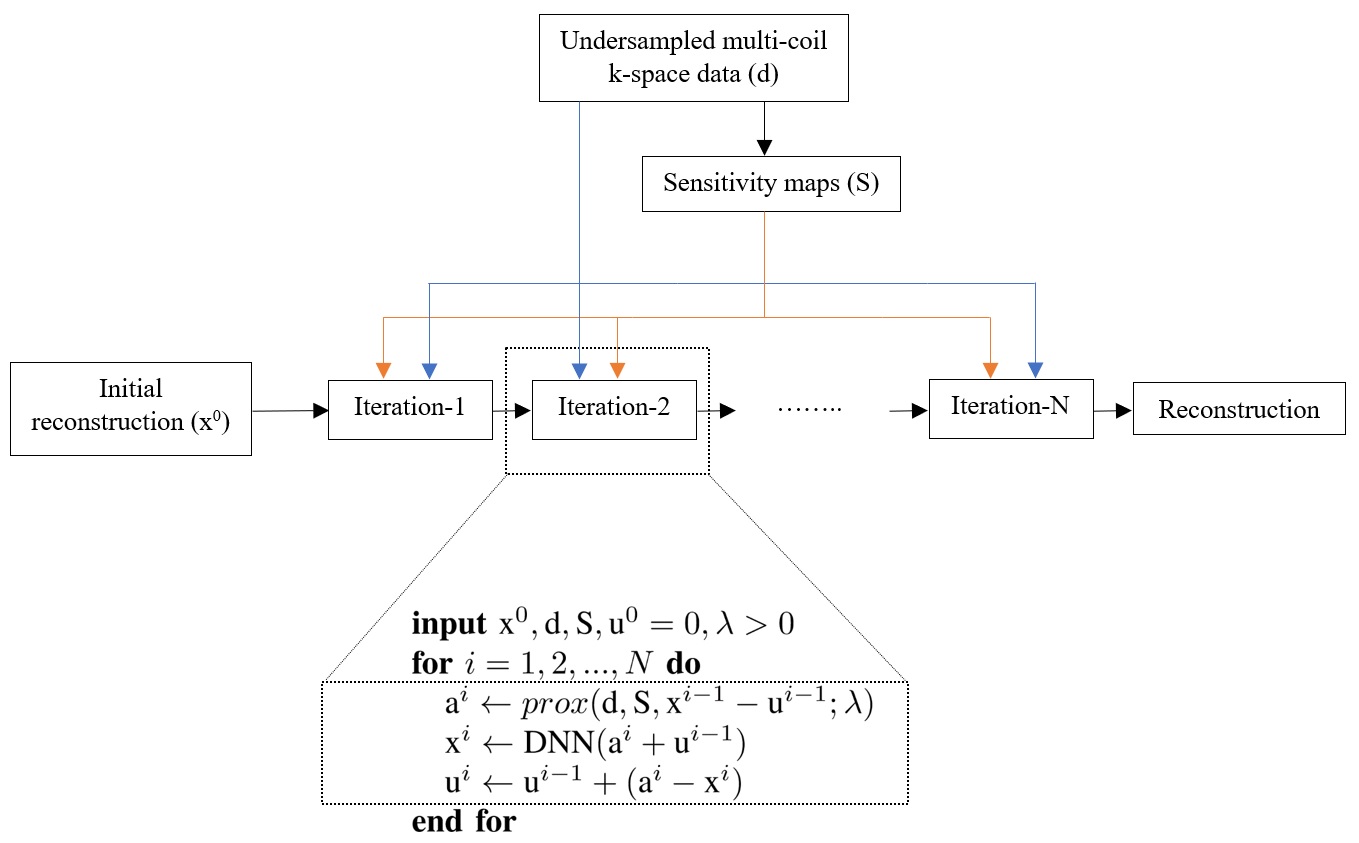}}
\end{center}
   \caption{Proposed deep plug-and-play prior framework.}
\label{fig1}
\end{figure*}

DNN architecture is an encoder-decoder Unet-type \cite{ronneberger2015u} convolutional network architecture with skip connections. The number of filters for both encoder and decoder layers is set to 128 and the network filter kernel size is set to 3 for both encoder and decoder layers. Mean-square-error (MSE) has been used as a Loss function and Loss minimization was performed using ADAM \cite{kingma2014adam} optimizer. Since the deep network frameworks work on real-valued parameters, inputs, and outputs, in our method complex data are divided into real and imaginary parts and considered as two-channel input and output. Figure \ref{fig1} illustrates the proposed ADMM based deep plug-and-play prior framework.
\\
\section{Results and Discussions}

In our experiments, we have tested our method with four different datasets in order to explore the generalization potential of our method for MRI reconstruction.\\

First dataset has been acquired (3D MPRAGE) on six volunteers with a total of 450 brain images used as the training set. A 32-channel head coil was used for the MPRAGE scans and the echo time (TE) of the scan was 2.17ms with a repetition time (TR) of 1.56s. We undersampled the multi-coil k-space data retrospectively with undersampling along both phase encoding dimensions (acceleration factor R=2x2). Written, informed consent was obtained from each volunteer prior to scanning and experiments were performed in accordance with the local IRB protocol.

For the second, third, and fourth datasets, we have used three knee datasets presented by \cite{hammernik2018learning}. Three datasets include Coronal PD dataset (knee dataset-1), Coronal fat-saturated proton-density (PD) dataset (knee dataset-2),  and Sagittal fat-saturated T2 dataset (knee dataset-3). In the knee datasets, each subjects were scanned with a 15-channel knee coil. 
Each of these three datasets includes a total of 200 images from 10 patients which have been used as the training set.  10 images from different patients for each dataset have used for testing purposes.

The sensitivity maps were computed from a block of size 24x24 for both brain and knee datasets using ESPIRiT \cite{uecker2014espirit} method. Full k-space data reconstructed with the adaptive combine method \cite{walsh2000adaptive} was used as our gold standard for comparison. 
The reconstruction performance was evaluated using quantitative metrics focusing on different aspects of the reconstruction quality. The Peak Signal to Noise Ratio (PSNR) was used to assess
the overall reconstructed image quality and the Structural Similarity Index (SSIM) was used to estimate the overall image similarity with respect to the reference reconstruction.

\begin{figure*}
\begin{center}
\centerline{\includegraphics[width=0.91\linewidth]{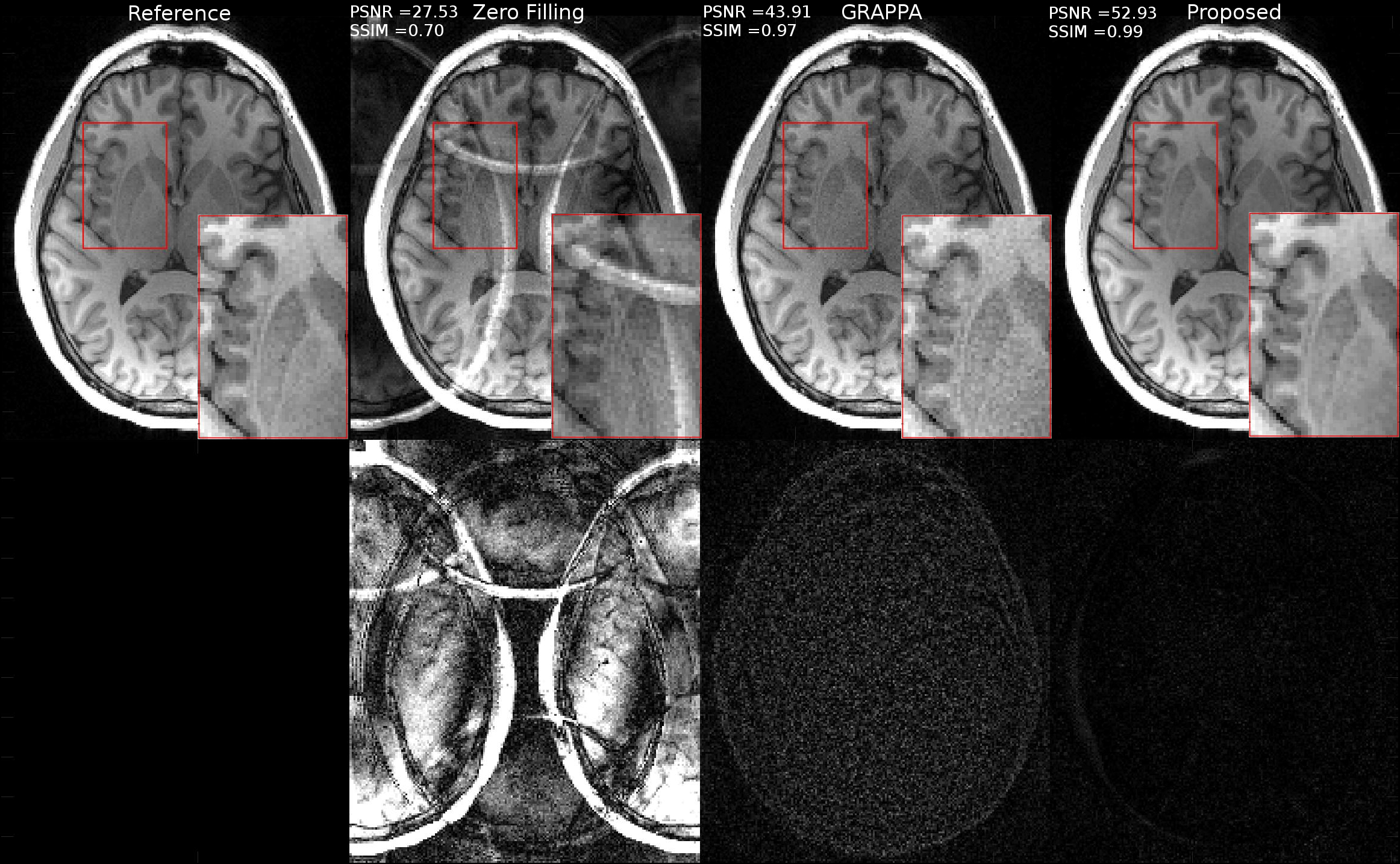}}
\end{center}
   \caption{First row (left to right): Gold standard reconstruction result using fully sampled data, zero-filled reconstruction, GRAPPA reconstruction result with undersampling factor of 2x2, and our reconstruction result with undersampling factor of 2x2 for 3D MPRAGE data. Second row, includes error maps correspond to each reconstruction results for comparison.}
\label{fig2}
\end{figure*}

\begin{figure*}
\begin{center}
\centerline{\includegraphics[width=\linewidth,scale=0.204]{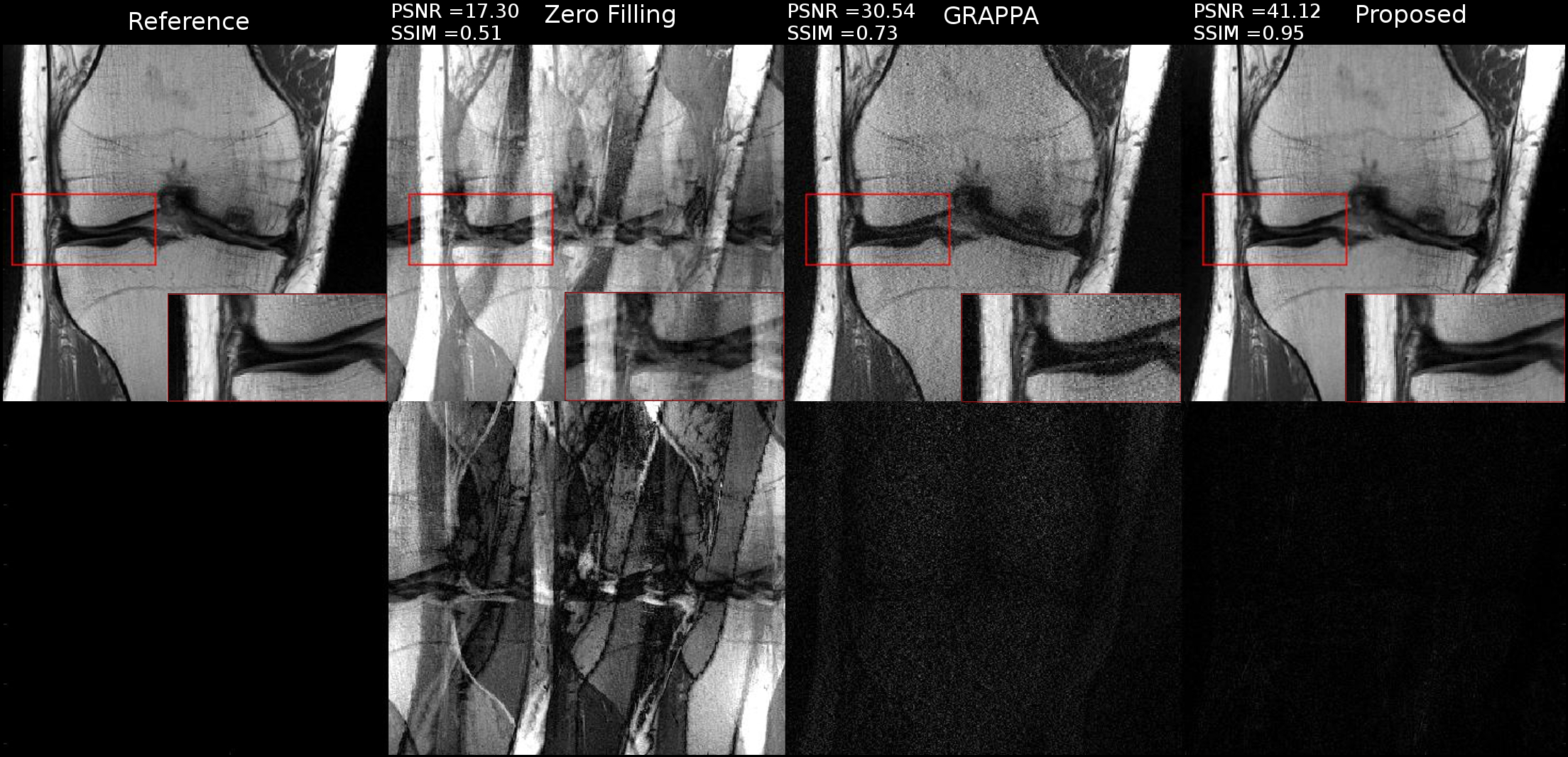}}
\end{center}

   \caption{First row (left to right): Gold standard reconstruction result using fully sampled data, zero-filled reconstruction, GRAPPA reconstruction result with undersampling factor of 4, and our reconstruction result with undersampling factor of 4 for 2D Coronal PD data. Second row, includes error maps correspond to each reconstruction results for comparison.}
\label{fig3}
\end{figure*}

\begin{figure*}
\begin{center}
\centerline{\includegraphics[width=\linewidth,scale=0.19]{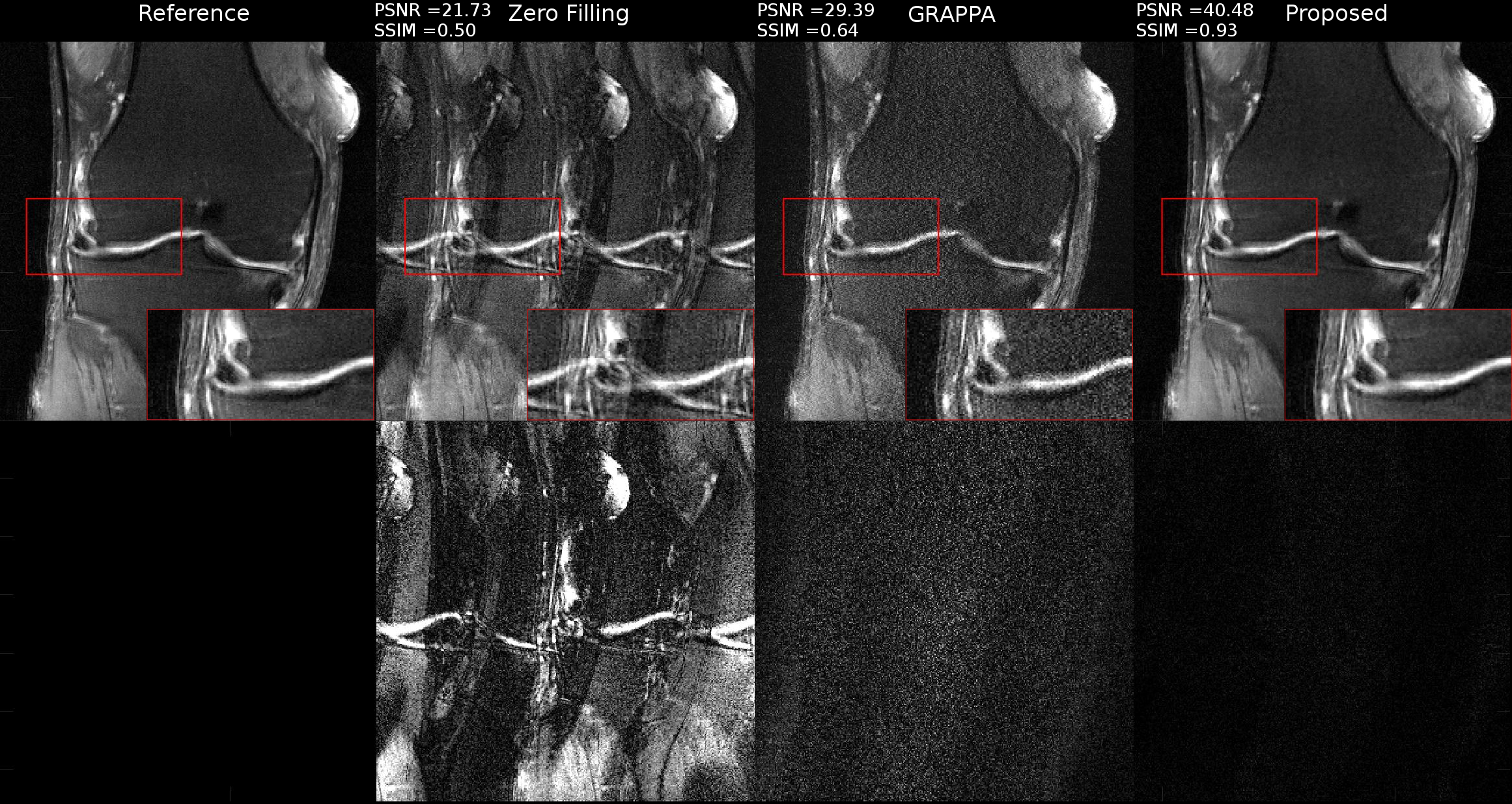}}
\end{center}
   \caption{First row (left to right): Gold standard reconstruction result using fully sampled data, zero-filled reconstruction, GRAPPA reconstruction result with undersampling factor of 4, and our reconstruction result with undersampling factor of 4 for 2D Coronal fat-saturated PD data. Second row, includes error maps correspond to each reconstruction results for comparison.}
\label{fig4}
\end{figure*}

Figure \ref{fig2} display the impact of acceleration factor R=2x2 for zero-filled reconstruction, the clinical gold standard GRAPPA, and our proposed method on 3D MPRAGE brain images (brain dataset). We observed that the proposed method reconstructs artifact-free images, which is sharper and have better quality than GRAPPA reconstruction, and GRAPPA result shows noise amplification compared to our result (PSNR of ours is 52.93 compared to PSNR of 43.91 for GRAPPA). Figures \ref{fig3},\ref{fig4} show the impact of acceleration factor R=4 (undersampling along only one phase encoding dimension) for zero-filled reconstruction, GRAPPA, and our proposed method on knee dataset-1 and knee dataset-2 respectively. Similar to Figure \ref{fig2}, GRAPPA results for knee data in Figures \ref{fig3} and \ref{fig4} show noise amplification compared to our results (PSNRs of ours are 41.12 and 40.48 compared to PSNRs of 30.54 and 29.39 for GRAPPA). Our results show better image restoration than GRAPPA and resulting in a nearly artifact-free reconstructed images. PSNR and SSIM quantitative variations on brain dataset is depicted in Table \ref{table1}. Table \ref{table2} shows PSNR and SSIM quantitative variations on knee datasets.
Tables \ref{table1},\ref{table2} show
that our reconstructions consistently have higher PSNRs and SSIMs than GRAPPA reconstructions.
\\

\begin{table}[h!]
\centering
\begin{tabular}{rllll}
\hline
\multicolumn{3}{c} {Brain Dataset} \\
\cline{2-3} 
\cline{4-5} 
Method & PSNR  & SSIM   \\
\hline
Proposed & $53.3\pm0.91$ & $0.99\pm0.0015$    \\
GRAPPA & $44.8\pm0.69$  & $0.97\pm0.0023$   \\

\hline
\end{tabular}
\caption{PSNR and SSIM variations on Brain dataset}
\label{table1}
\end{table}

\begin{table*}[h!]
\centering
\begin{tabular}{rlllllll}
\hline
\multicolumn{3}{c} {Knee Dataset-1} & {Knee Dataset-2} & & {Knee Dataset-3} \\
\cline{2-3} 
\cline{4-5} 
\cline{6-7}
Method & PSNR & SSIM & PSNR & SSIM & PSNR & SSIM \\
\hline
Proposed & $42.28\pm1.31$ & $0.96\pm0.0108$  & $39.87\pm1.08$ & $0.93\pm0.0086$ & $44.09\pm2.85$ & $0.95\pm0.053$ \\
GRAPPA & $30.27\pm0.89$ & $0.71\pm0.081$ & $28.43\pm0.97$ & $0.6\pm0.045$ & $31.91\pm1.54$ & $0.7\pm0.063$ \\

\hline
\end{tabular}
\caption{PSNR and SSIM variations on Knee datasets-1,2,3}
\label{table2}
\end{table*}

\section{Conclusion}
This paper proposes an ADMM based deep plug-and-play prior framework and demonstrates the effectiveness of learning-based plug-and-play prior framework for parallel MRI reconstruction. The reported results on four real (not simulated) MRI datasets show that our proposed method outperforms the clinical gold standard GRAPPA method. We have demonstrated that the image quality arising from partially parallel MRI reconstruction can be improved, in comparison to the GRAPPA reconstruction, by using the proposed ADMM based deep plug-and-play prior framework.

\section*{Acknowledgment}
This research was supported in part by NIH grants R01 NS079788, R01 EB019483, R01 DK100404, R44 MH086984, IDDRC U54 HD090255, and by a research grant from the Boston Children's Hospital Translational Research Program.

{\small
\bibliographystyle{ieee}
\bibliography{egpaper_final}

\begin{thebibliography}{10}\itemsep=-1pt

\bibitem{bammer2007augmented}
R.~Bammer, M.~Aksoy, and C.~Liu.
\newblock Augmented generalized sense reconstruction to correct for rigid body
  motion.
\newblock {\em Magnetic Resonance in Medicine}, 57(1):90--102, 2007.

\bibitem{bilgic2015wave}
B.~Bilgic, B.~A. Gagoski, S.~F. Cauley, A.~P. Fan, J.~R. Polimeni, P.~E. Grant,
  L.~L. Wald, and K.~Setsompop.
\newblock Wave-caipi for highly accelerated 3d imaging.
\newblock {\em Magnetic resonance in medicine}, 73(6):2152--2162, 2015.

\bibitem{boyd2011distributed}
S.~Boyd, N.~Parikh, E.~Chu, B.~Peleato, J.~Eckstein, et~al.
\newblock Distributed optimization and statistical learning via the alternating
  direction method of multipliers.
\newblock {\em Foundations and Trends{\textregistered} in Machine learning},
  3(1):1--122, 2011.

\bibitem{breuer2006controlled}
F.~A. Breuer, M.~Blaimer, M.~F. Mueller, N.~Seiberlich, R.~M. Heidemann, M.~A.
  Griswold, and P.~M. Jakob.
\newblock Controlled aliasing in volumetric parallel imaging (2d caipirinha).
\newblock {\em Magnetic resonance in medicine}, 55(3):549--556, 2006.

\bibitem{chan2017plug}
S.~H. Chan, X.~Wang, and O.~A. Elgendy.
\newblock Plug-and-play admm for image restoration: Fixed-point convergence and
  applications.
\newblock {\em IEEE Transactions on Computational Imaging}, 3(1):84--98, 2017.

\bibitem{griswold2002generalized}
M.~A. Griswold, P.~M. Jakob, R.~M. Heidemann, M.~Nittka, V.~Jellus, J.~Wang,
  B.~Kiefer, and A.~Haase.
\newblock Generalized autocalibrating partially parallel acquisitions (grappa).
\newblock {\em Magnetic resonance in medicine}, 47(6):1202--1210, 2002.

\bibitem{griswold2000partially}
M.~A. Griswold, P.~M. Jakob, M.~Nittka, J.~W. Goldfarb, and A.~Haase.
\newblock Partially parallel imaging with localized sensitivities (pils).
\newblock {\em Magnetic Resonance in Medicine}, 44(4):602--609, 2000.

\bibitem{guerquin2012realistic}
M.~Guerquin-Kern, L.~Lejeune, K.~P. Pruessmann, and M.~Unser.
\newblock Realistic analytical phantoms for parallel magnetic resonance
  imaging.
\newblock {\em IEEE Transactions on Medical Imaging}, 31(3):626--636, 2012.

\bibitem{haldar2014low}
J.~P. Haldar.
\newblock Low-rank modeling of local $ k $-space neighborhoods (loraks) for
  constrained mri.
\newblock {\em IEEE transactions on medical imaging}, 33(3):668--681, 2014.

\bibitem{hammernik2018learning}
K.~Hammernik, T.~Klatzer, E.~Kobler, M.~P. Recht, D.~K. Sodickson, T.~Pock, and
  F.~Knoll.
\newblock Learning a variational network for reconstruction of accelerated mri
  data.
\newblock {\em Magnetic resonance in medicine}, 79(6):3055--3071, 2018.

\bibitem{heberlein2006auto}
K.~Heberlein and X.~Hu.
\newblock Auto-calibrated parallel spiral imaging.
\newblock {\em Magnetic resonance in medicine}, 55(3):619--625, 2006.

\bibitem{heidemann2001vd}
R.~M. Heidemann, M.~A. Griswold, A.~Haase, and P.~M. Jakob.
\newblock Vd-auto-smash imaging.
\newblock {\em Magnetic resonance in medicine}, 45(6):1066--1074, 2001.

\bibitem{jakob1998auto}
P.~M. Jakob, M.~A. Griswold, R.~R. Edelman, and D.~K. Sodickson.
\newblock Auto-smash: a self-calibrating technique for smash imaging.
\newblock {\em Magnetic Resonance Materials in Physics, Biology and Medicine},
  7(1):42--54, 1998.

\bibitem{kamilov2017plug}
U.~S. Kamilov, H.~Mansour, and B.~Wohlberg.
\newblock A plug-and-play priors approach for solving nonlinear imaging inverse
  problems.
\newblock {\em IEEE Signal Processing Letters}, 24(12):1872--1876, 2017.

\bibitem{kingma2014adam}
D.~P. Kingma and J.~Ba.
\newblock Adam: A method for stochastic optimization.
\newblock {\em arXiv preprint arXiv:1412.6980}, 2014.

\bibitem{kyriakos2000sensitivity}
W.~E. Kyriakos, L.~P. Panych, D.~F. Kacher, C.-F. Westin, S.~M. Bao, R.~V.
  Mulkern, and F.~A. Jolesz.
\newblock Sensitivity profiles from an array of coils for encoding and
  reconstruction in parallel (space rip).
\newblock {\em Magnetic Resonance in Medicine}, 44(2):301--308, 2000.

\bibitem{liang2009accelerating}
D.~Liang, B.~Liu, J.~Wang, and L.~Ying.
\newblock Accelerating sense using compressed sensing.
\newblock {\em Magnetic Resonance in Medicine}, 62(6):1574--1584, 2009.

\bibitem{liu2007parallel}
C.~Liu, R.~Bammer, and M.~E. Moseley.
\newblock Parallel imaging reconstruction for arbitrary trajectories using
  k-space sparse matrices (kspa).
\newblock {\em Magnetic resonance in medicine}, 58(6):1171--1181, 2007.

\bibitem{lustig2007sparse}
M.~Lustig, D.~Donoho, and J.~M. Pauly.
\newblock Sparse mri: The application of compressed sensing for rapid mr
  imaging.
\newblock {\em Magnetic resonance in medicine}, 58(6):1182--1195, 2007.

\bibitem{lustig2010spirit}
M.~Lustig and J.~M. Pauly.
\newblock Spirit: Iterative self-consistent parallel imaging reconstruction
  from arbitrary k-space.
\newblock {\em Magnetic resonance in medicine}, 64(2):457--471, 2010.

\bibitem{pruessmann2001advances}
K.~P. Pruessmann, M.~Weiger, P.~B{\"o}rnert, and P.~Boesiger.
\newblock Advances in sensitivity encoding with arbitrary k-space trajectories.
\newblock {\em Magnetic resonance in medicine}, 46(4):638--651, 2001.

\bibitem{pruessmann1999sense}
K.~P. Pruessmann, M.~Weiger, M.~B. Scheidegger, P.~Boesiger, et~al.
\newblock Sense: sensitivity encoding for fast mri.
\newblock {\em Magnetic resonance in medicine}, 42(5):952--962, 1999.

\bibitem{ramani2011parallel}
S.~Ramani and J.~A. Fessler.
\newblock Parallel mr image reconstruction using augmented lagrangian methods.
\newblock {\em IEEE Transactions on Medical Imaging}, 30(3):694--706, 2011.

\bibitem{ronneberger2015u}
O.~Ronneberger, P.~Fischer, and T.~Brox.
\newblock U-net: Convolutional networks for biomedical image segmentation.
\newblock In {\em International Conference on Medical image computing and
  computer-assisted intervention}, pages 234--241. Springer, 2015.

\bibitem{seiberlich2008reconstruction}
N.~Seiberlich, F.~Breuer, R.~Heidemann, M.~Blaimer, M.~Griswold, and P.~Jakob.
\newblock Reconstruction of undersampled non-cartesian data sets using
  pseudo-cartesian grappa in conjunction with grog.
\newblock {\em Magnetic resonance in medicine}, 59(5):1127--1137, 2008.

\bibitem{sodickson1997simultaneous}
D.~K. Sodickson and W.~J. Manning.
\newblock Simultaneous acquisition of spatial harmonics (smash): fast imaging
  with radiofrequency coil arrays.
\newblock {\em Magnetic resonance in medicine}, 38(4):591--603, 1997.

\bibitem{sreehari2016plug}
S.~Sreehari, S.~V. Venkatakrishnan, B.~Wohlberg, G.~T. Buzzard, L.~F. Drummy,
  J.~P. Simmons, and C.~A. Bouman.
\newblock Plug-and-play priors for bright field electron tomography and sparse
  interpolation.
\newblock {\em IEEE Transactions on Computational Imaging}, 2(4):408--423,
  2016.

\bibitem{sun2019online}
Y.~Sun, B.~Wohlberg, and U.~S. Kamilov.
\newblock An online plug-and-play algorithm for regularized image
  reconstruction.
\newblock {\em IEEE Transactions on Computational Imaging}, 2019.

\bibitem{uecker2014espirit}
M.~Uecker, P.~Lai, M.~J. Murphy, P.~Virtue, M.~Elad, J.~M. Pauly, S.~S.
  Vasanawala, and M.~Lustig.
\newblock Espirit—an eigenvalue approach to autocalibrating parallel mri:
  where sense meets grappa.
\newblock {\em Magnetic resonance in medicine}, 71(3):990--1001, 2014.

\bibitem{venkatakrishnan2013plug}
S.~V. Venkatakrishnan, C.~A. Bouman, and B.~Wohlberg.
\newblock Plug-and-play priors for model based reconstruction.
\newblock In {\em 2013 IEEE Global Conference on Signal and Information
  Processing}, pages 945--948. IEEE, 2013.

\bibitem{walsh2000adaptive}
D.~O. Walsh, A.~F. Gmitro, and M.~W. Marcellin.
\newblock Adaptive reconstruction of phased array mr imagery.
\newblock {\em Magnetic Resonance in Medicine}, 43(5):682--690, 2000.

\bibitem{yazdanpanah2017compressed}
A.~P. Yazdanpanah and E.~E. Regentova.
\newblock Compressed sensing magnetic resonance imaging based on shearlet
  sparsity and nonlocal total variation.
\newblock {\em Journal of Medical Imaging}, 4(2):026003, 2017.

\bibitem{ayazdanpanah2017}
A.~P. Yazdanpanah and E.~E. Regentova.
\newblock Compressed sensing mri using curvelet sparsity and nonlocal total
  variation: Cs-nltv.
\newblock {\em Electronic Imaging}, 2017(13):5--9, 2017.

\bibitem{yeh20053parallel}
E.~N. Yeh, C.~A. McKenzie, M.~A. Ohliger, and D.~K. Sodickson.
\newblock 3parallel magnetic resonance imaging with adaptive radius in k-space
  (pars): Constrained image reconstruction using k-space locality in
  radiofrequency coil encoded data.
\newblock {\em Magnetic resonance in medicine}, 53(6):1383--1392, 2005.

\end{thebibliography}
}

\end{document}